\documentclass{article}
\usepackage{spconf,amsmath,graphicx,hyperref,enumitem,booktabs,multirow}
\usepackage[caption=false,font=footnotesize]{subfig}

\title{SCENE: Semantic-aware Codec Enhancement with Neural Embeddings}
\name{Han-Yu Lin$^{1}$
\qquad Li-Wei Chen$^{1,2}$
\qquad Hung-Shin Lee$^{1}$}

\address{$^{1}$ United Link Co., Ltd., Taiwan \\
$^{2}$ Dept. Computer Science and Information Engineering, National Tsing Hua University, Taiwan}

\begin{document}
%
\maketitle
\begin{abstract}
Compression artifacts from standard video codecs often degrade perceptual quality.
We propose a lightweight, semantic-aware pre-processing framework that enhances perceptual fidelity by selectively addressing these distortions.
Our method integrates semantic embeddings from a vision-language model into an efficient convolutional architecture, prioritizing the preservation of perceptually significant structures.
The model is trained end-to-end with a differentiable codec proxy, enabling it to mitigate artifacts from various standard codecs without modifying the existing video pipeline.
During inference, the codec proxy is discarded, and SCENE operates as a standalone pre-processor, enabling real-time performance.
Experiments on high-resolution benchmarks show improved performance over baselines in both objective (MS-SSIM) and perceptual (VMAF) metrics, with notable gains in preserving detailed textures within salient regions.
Our results show that semantic-guided, codec-aware pre-processing is an effective approach for enhancing compressed video streams.
\end{abstract}

\begin{figure*}[ht]
\centering
\includegraphics[width=1.0\linewidth]{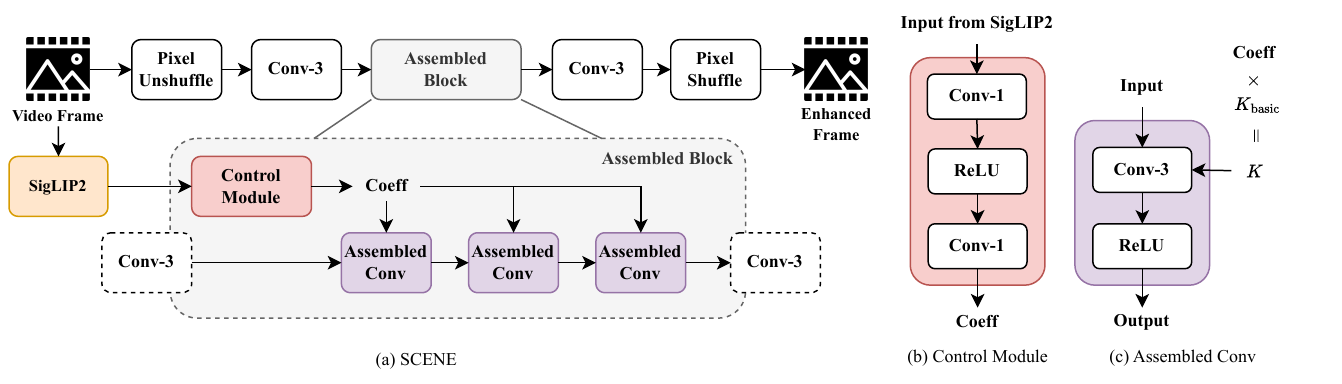}
\vspace{-25pt}
\caption{Illustration of the proposed SCENE framework.}
\label{fig:main}
\vspace{-10pt}
\end{figure*}
\section{Introduction}
\label{sec:intro}

The rapid growth of high-resolution video, from 4K streaming to immersive media, has created a fundamental tension between user expectations for perceptual quality and the persistent constraints of network bandwidth.
Standard video codecs such as H.264 (AVC) and H.265 (HEVC) require aggressive compression, often resulting in visible artifacts.
This disparity between compression efficiency and human perception has motivated significant research in neural codec enhancement, aiming to restore visual fidelity at low bitrates.

Prior work in this area can be categorized along several axes.
Post-encoding methods enhance quality on the decoder side through artifact removal or diffusion-based restoration~\cite{huang2022, zhou2025}, with some systems providing Quality of Experience (QoE) guarantees~\cite{wang2024}.
Pre-encoding methods apply learned preprocessing to optimize content for standard encoders, achieving substantial bitrate savings~\cite{chadha2021, zhao2024}.
End-to-end neural codecs replace the traditional pipeline entirely and have shown gains over next-generation codecs such as AV1 and VVC~\cite{jia2025, khan2025}.
Finally, existing semantic-aware methods rely on simple saliency maps and do not exploit rich object-level context~\cite{bie2025, yang2024}.

Despite these advances, a key perceptual gap remains.
Current methods are largely semantically agnostic, optimizing for pixel-level metrics (PSNR/SSIM), thus often restoring background textures at the expense of perceptually critical regions such as faces or text.
This limitation is compounded by a training-deployment mismatch, as models trained on idealized codec proxies struggle to generalize to the complex distortions introduced by real-world codecs.

Vision-language models (VLMs) such as CLIP and BLIP-2~\cite{radford2021, li2023} offer a promising direction to bridge this semantic gap.
Their embeddings encode rich semantic information, yet their application to video compression remains underexplored.
We propose \textbf{SCENE (Semantic-aware Codec Enhancement with Neural Embeddings)}, a lightweight framework that integrates VLM-derived semantic guidance with efficient assembled convolutions.
To address the codec mismatch, we train with a differentiable proxy so that at inference, only the enhancement pre-processor is needed for integration into existing pipelines.

Our contributions are as follows:
\begin{enumerate}[noitemsep,leftmargin=*,topsep=1pt]
\item We introduce SCENE, a semantic-aware enhancement framework that combines two key components.
First, it leverages VLM embeddings with assembled convolutions to prioritize the restoration of perceptually critical content.
Second, it employs a training strategy based on a differentiable codec proxy, bridging the training-deployment gap by aligning the model with real-world codec distortions.
\item Experiments on high-resolution benchmarks show that SCENE achieves consistent improvements over a super-resolution baseline~\cite{guo2023} in perceptual quality metrics.
\end{enumerate}

\section{Methodology}
\label{sec:format}
Fig.~\ref{fig:main} illustrates SCENE's architecture.
Input frames are first downsampled via pixel-unshuffle with scale factor $N{=}2$ \cite{guo2023}, downscaling spatial resolution by a factor of $N^2{=}4$ while increasing channel depth to lower computational cost.
A $3\times3$ convolutional layer extracts low-level features from the downsampled input.
In parallel, a frozen SigLIP 2 encoder \cite{tschannen2025} extracts semantic embeddings that are transformed into convolution coefficients via a control module.
These coefficients modulate assembled blocks for content-adaptive enhancement.
After a final $3\times3$ convolution, pixel-shuffle reconstructs the enhanced frame at the original resolution.
During training, a differentiable codec proxy simulates compression artifacts.
At inference, the codec proxy is removed, and SCENE operates solely as a standalone pre-processor.

\subsection{Semantic Feature Extraction}
SigLIP 2 So400M extracts 1,152-dimensional semantic embeddings from input frames.
Unlike its predecessor, which focuses primarily on global alignment, SigLIP 2 is optimized for dense predictions and localization via a unified training recipe involving decoder-based losses and self-distillation.
This ensures that the extracted embeddings $F_{\text{sem}}$ not only encode high-level concepts like objects and scenes but also preserve fine-grained spatial structures and texture details.

These spatially-aware semantic features enable SCENE to prioritize perceptually meaningful structures effectively.
The embeddings are transformed into channel-specific convolution coefficients via a control module (Fig.~\ref{fig:main}(b)).
The control module applies two $1\times1$ convolutions with ReLU activation:
\begin{equation}
\text{Coeff} = \text{Conv}_{1\times1}(\text{ReLU}(\text{Conv}_{1\times1}(F_{\text{sem}})))
\end{equation}
The resulting coefficients modulate the assembled convolution kernels, enabling content-adaptive enhancement guided by semantic priors.

\subsection{Assembled Convolutions}
Fig.~\ref{fig:main}(c) illustrates our assembled convolution mechanism.
Unlike standard convolutions with fixed kernels, we dynamically construct kernels from a set of base kernels $\{k_i\}$ using channel-specific coefficients predicted by the control module in Fig.~\ref{fig:main}(b):
\begin{equation}
\label{eq:kc}
K_c = \sum_{i=1}^{E} \text{Coeff}_{c,i} \cdot k_i
\end{equation}
In Eq.~\ref{eq:kc}, $K_c$ is the assembled kernel for output channel $c$, and $E{=}4$ is the number of base kernels.
This formulation differs from dynamic convolution, which applies the same coefficients across all channels; assembled convolution computes independent coefficients for each output channel, providing finer-grained adaptivity.

As shown in Fig.~\ref{fig:main}(a), SCENE employs two assembled blocks, each containing three assembled convolutions with 64 channels.
The first block is modulated by SigLIP 2 features, while the second block derives its coefficients from the preceding convolutional feature maps, enabling progressive semantic-to-spatial refinement.
The assembled kernels enable content-adaptive filtering, allocating more representational power to perceptually important regions while maintaining computational efficiency.

\subsection{Training Strategy}
During training, we employ a differentiable JPEG proxy that captures block-based transform–quantization distortions common
to H.264 and H.265 \cite{guleryuz2025}, while omitting motion-compensated prediction,
thereby enabling end-to-end optimization.
Enhanced frames pass through this proxy before loss computation. The multi-stage loss function is
\begin{equation}
\label{eq:loss}
L_{\text{total}} = \lambda_p L_{\text{perceptual}} + \lambda_b L_{\text{bitrate}} + \lambda_1 L_{\text{pre}} + \lambda_2 L_{\text{post}}.
\end{equation}

In Eq.~\ref{eq:loss}, $L_{\text{perceptual}}$ is computed using a differentiable PyTorch implementation of VMAF~\cite{aistov2023,vmaf2018}, enabling gradient-based optimization of perceptual quality.
$L_{\text{bitrate}}$ estimates bandwidth from quantized coefficients.
$L_{\text{pre}}$ and $L_{\text{post}}$ are L1 reconstruction losses computed between the input frame and the enhanced frame before and after the JPEG proxy, respectively, providing stable gradients when perceptual losses are noisy.
During inference, only the pre-processor is active, operating without the JPEG proxy.
This asymmetric design enables learning compression-robust enhancements while preserving real-time inference.

\begin{table}[t]
\centering
\caption{BD-rate (\%) comparison of our proposed SCENE against the AsConvSR baseline across standard codecs. Negative values indicate bitrate savings.}
\vspace{5pt}
\label{tab:results}
\resizebox{\columnwidth}{!}{
\begin{tabular}{lcccc}
\toprule
Codec & Model & MS-SSIM ↓ & VMAF ↓ & VMAF\_NEG ↓ \\
\midrule\midrule
\multirow{2}{*}{H.264} & AsConvSR & 6.1 & -29.4 & -11.9 \\
& \textbf{SCENE} & 6.7 & -32.0 & -12.0 \\
\midrule
\multirow{2}{*}{H.265} & AsConvSR & 11.5 & -33.9 & -7.9 \\
& \textbf{SCENE} & 11.0 & -37.4 & -8.6 \\
\midrule
\multirow{2}{*}{AV1} & AsConvSR & 14.65 & N/A* & -5.23\\
& \textbf{SCENE} & 15.02 & N/A* & -5.39\\
\bottomrule
\end{tabular}}
\vspace{-5pt}
\footnotesize{\textsuperscript{*}No overlapping VMAF interval.}
\end{table}

\section{Experiments}
\label{sec:pagestyle}

\subsection{Dataset and Preprocessing}

SCENE was trained on the Vimeo-90K dataset \cite{xue2019}, a widely adopted benchmark containing diverse short video clips for video enhancement.
A 9:1 train/validation split was used, adhering to standard practices.
Input frames were randomly cropped to $256 \times 256$ patches with random flips for data augmentation.
We evaluated on the UVG 1080p test set \cite{mercat2020}, a standard benchmark for high-resolution streaming.

\subsection{Implementation Details}
A frozen SigLIP 2 So400M backbone \cite{tschannen2025} was employed for semantic feature extraction, combined with two assembled blocks \cite{guo2023} as described in Section~\ref{sec:format}.
Each block consisted of three convolutional layers with 64 channels, kernel size $3 \times 3$, and $E{=}4$ base kernels.
The differentiable JPEG proxy \cite{guleryuz2025} was used to simulate compression artifacts during training.
The model had 1.4M trainable parameters.

We trained using schedule-free AdamW with an initial learning rate of $1\times10^{-3}$ for 30 epochs \cite{defazio2024}.
The multi-stage loss function was defined in Eq.~\ref{eq:loss}, with weights set as $\lambda_p=1\times10^{-2}$, $\lambda_b=1$, $\lambda_1=5$, and $\lambda_2=1$.
We used a batch size of 128 per GPU across four NVIDIA L40S units, with a total training time of approximately five hours.

\subsection{Evaluation Protocol}
We evaluated rate–distortion performance using BD-rate \cite{gisle2008} computed over VMAF, VMAF\_NEG, and MS-SSIM \cite{wang2003}.
VMAF is a perceptual video quality assessment algorithm developed by Netflix \cite{vmaf2018}; VMAF\_NEG extends it with a term penalizing unnatural enhancements such as sharpening artifacts.
MS-SSIM captures multi-scale structural fidelity.
All metrics were computed on the UVG dataset for H.264, H.265, and AV1 codecs.
Inference latency was measured at 1080p resolution to assess deployment feasibility.

\begin{figure}[t]
\centering
\includegraphics[width=0.45\textwidth]{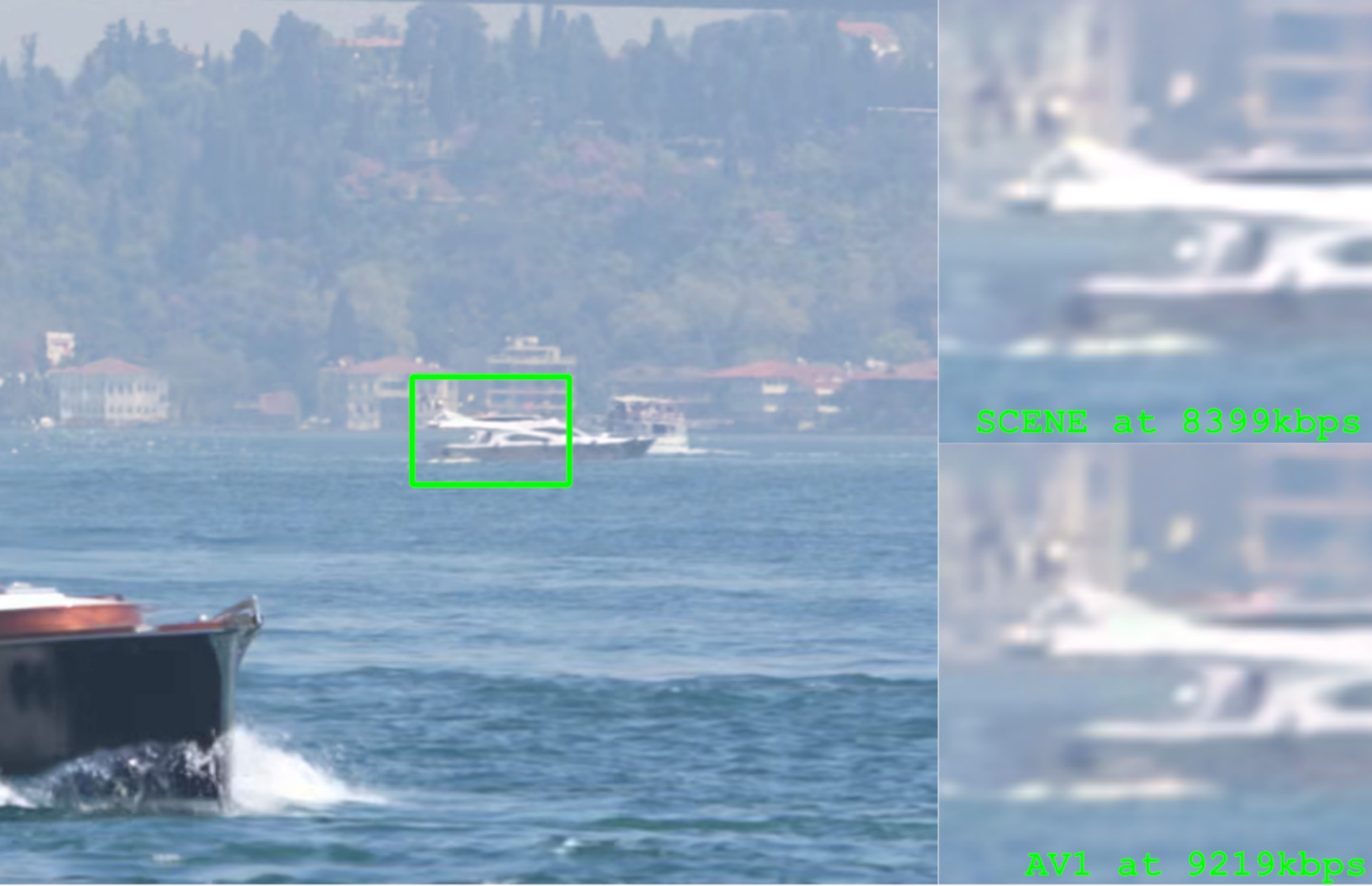}
\vspace{-10pt}
\caption{Qualitative comparison between standard AV1 and our SCENE-enhanced AV1.}
\label{fig:qualitative}
\vspace{-10pt}
\end{figure}

\subsection{Baseline Methods}
We compared SCENE against two baselines: 1) \textbf{AsConvSR} \cite{guo2023}: an assembled convolution network sharing the same architecture as SCENE, but without semantic modulation.
Both assembled blocks derive coefficients from convolutional feature maps rather than VLM embeddings, serving as an ablation of semantic guidance.
2) \textbf{Codec only}: standard H.264, H.265, and AV1 compression without any pre-processing.
These comparisons isolate the contributions of semantic guidance and codec-aware training.

\begin{figure*}[ht]
\centering
\includegraphics[width=1\textwidth]{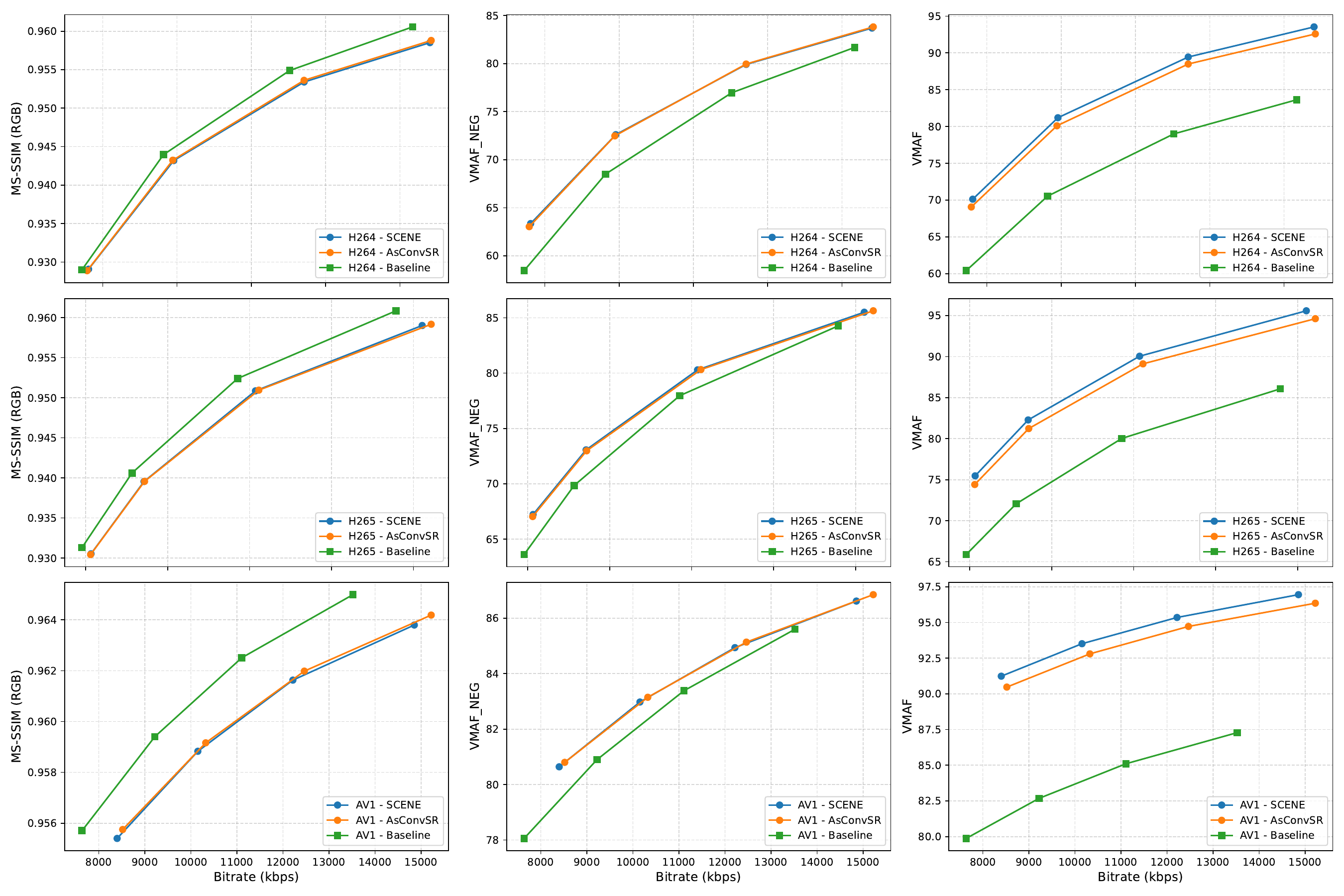}
\vspace{-25pt}
\caption{Rate-distortion curves for H.264, H.265, and AV1 on MS-SSIM, VMAF\textunderscore NEG, and VMAF, respectively.}
\label{fig:rate-distortion}
\vspace{-10pt}
\end{figure*}

\subsection{Results and Analysis}

Table~\ref{tab:results} summarizes the BD-rate results of SCENE and the AsConvSR baseline for H.264, H.265, and AV1, each evaluated against the codec-only anchor.
For H.264, SCENE achieved a VMAF BD-rate reduction of 32.0\%, compared with 29.4\% for AsConvSR.
In a direct comparison between the two enhancement models, SCENE reduced BD-rate by 3.9\% relative to AsConvSR, indicating that semantic guidance yielded additional rate-quality gains beyond spatially adaptive assembled convolutions.
For H.265, the VMAF BD-rate reductions were 37.4\% for SCENE and 33.9\% for AsConvSR; the direct comparison indicated a 5.8\% reduction in favor of SCENE.
In contrast, MS-SSIM BD-rate values were slightly positive (+6--11\%) for both methods, suggesting that optimizing for perceptual metrics (VMAF/VMAF\_NEG) does not necessarily improve pixel-level fidelity.
Notably, SCENE remained within 0.5\% of AsConvSR on MS-SSIM, implying that semantic modulation improved perceptual quality without materially degrading structural similarity relative to the baseline.

For AV1, both enhancement models improved perceptual quality, with SCENE achieving VMAF gains of up to 10.6 points.
However, the enhancement-induced bitrate increase shifted the enhanced rate-distortion curves outside the codec-only operating range, yielding no overlapping VMAF interval and therefore rendering BD-rate undefined (Table~\ref{tab:results}).
This behavior differed from that of H.264/H.265, where enhancement yielded bitrate reductions, and might have been attributable to interactions between AV1's in-loop tools and psychovisual optimizations, as well as the priors introduced by neural pre-processing.
Given that AV1 deployment is less prevalent in our target scenario, we primarily focused on H.264 and H.265, where SCENE exhibited consistent and interpretable gains under codec distortions.

Qualitative comparisons in Fig.~\ref{fig:qualitative} support the quantitative results.
At low bitrates, where conventional encoding introduced noticeable blocking, blurring, and texture attenuation, SCENE better preserved salient object boundaries and fine structures, as evident in the enhanced yacht where finer hull and edge details were retained.

The rate-distortion curves in Fig.~\ref{fig:rate-distortion} further corroborate these trends: SCENE consistently improved VMAF by 1--2 points over AsConvSR and by 10--12 points over the codec-only anchor, with larger gains observed in lower-bitrate regimes.
Similar improvements were observed for VMAF\_NEG, suggesting that the perceptual gains were not driven by unnatural sharpening artifacts.

Overall, the performance gap between SCENE and AsConvSR was modest, which was expected given that AsConvSR already provided strong restoration capability through spatially adaptive assembled convolutions.
Moreover, only the first assembled block was conditioned on VLM-derived semantic features, whereas the second block relied on learned convolutional priors.
Consequently, SCENE primarily refined perceptually salient regions while preserving structural fidelity, with the most pronounced gains observed in lower-bitrate regimes.

\section{Conclusion and Future Work}
\label{sec:typestyle}
We present \textbf{SCENE}, a lightweight semantic-guided, codec-aware enhancement framework that can be inserted before standard codecs.
SCENE conditions assembled convolutions on frozen VLM embeddings (SigLIP 2) and is trained with a differentiable codec proxy to reduce the training-deployment gap.
On UVG, SCENE yielded consistent VMAF/VMAF\_NEG BD-rate reductions over AsConvSR under H.264 and H.265, while maintaining comparable MS-SSIM.
SCENE has 1.4M trainable parameters and an inference latency of 27.74~ms per 1080p frame ($\sim$36~fps) on an RTX~4090, indicating suitability for real-time deployment.

Our future work will strengthen semantic conditioning, incorporate temporal modeling, and extend support to additional codecs and streaming settings.

\bibliographystyle{IEEEbib}
\bibliography{references}

\end{document}